\documentclass[12pt,preprint]{aastex}

\begin{document}

\title{The Design of Radio Telescope Array Configurations
using Multiobjective Optimization: \\ Imaging Performance versus
Cable Length}


\author{Babak E. Cohanim\altaffilmark{1}}
\affil{Center for Space Research, MIT,
    Cambridge, MA 02139}

\author{Jacqueline N. Hewitt\altaffilmark{2}}
\affil{Center for Space Research, MIT,
    Cambridge, MA 02139}

\and

\author{Olivier de Weck\altaffilmark{3}}
\affil{Space Systems Laboratory, MIT,
    Cambridge, MA 02139}


\altaffiltext{1}{Research Assistant, Aeronautics/Astronautics}
\altaffiltext{2}{Professor of Physics} \altaffiltext{3}{Assistant
Professor of Aeronautics/Astronautics and Engineering Systems}


\begin{abstract}
The next generation of radio telescope interferometric arrays
requires careful design of the array configuration to optimize the
performance of the overall system.  We have developed a framework,
based on a genetic algorithm, for rapid exploration and
optimization of the objective space pertaining to multiple
objectives.  We have evaluated a large space of possible designs
for 27-, 60-, 100-, and 160-station arrays.  The 27-station
optimizations can be compared to the well-known VLA case, and the
larger array designs apply to arrays currently under design such
as LOFAR, ATA, and the SKA.  In the initial implementation of our
framework we evaluate designs with respect to two metrics, array
imaging performance and the length of cable necessary to connect
the stations.  Imaging performance is measured by the degree to
which the sampling of the $uv$ plane is uniform.  For the larger
arrays we find that well-known geometric designs perform well and
occupy the Pareto front of optimum solutions.  For the 27-element
case we find designs, combining features of the well-known
designs, that are more optimal as measured by these two metrics.
The results obtained by the multiobjective genetic optimization
are corroborated by simulated annealing, which also reveals the
role of entropy in array optimization. Our framework is general,
and may be applied to other design goals and issues, such as
particular schemes for sampling the $uv$ plane, array robustness,
and phased deployment of arrays.
\end{abstract}




\section{Introduction}

A central issue in the design of a radio astronomical correlating
array is its configuration.  The placement of the antennas
determines the sampling of the Fourier transform of the sky
brightness distribution \citep{Thompson:1986} and hence the
fidelity of the image computed from the interferometric data.  The
placement of the antennas also affects the cost of the array by
determining the costs of power and signal distribution, site
preparation, roads, and other infrastructure items.  Typically
these considerations lead to trade-offs that must be considered in
the design.  Antenna arrays use earth rotation aperture synthesis
to sample the $uv$ plane over time as the earth rotates.  For
arrays which are required to operate over a wide range of
declinations and for arrays where instantaneous capabilities are
important, two-dimensional array configurations must be
considered. Two-dimensional arrays offer the instantaneous $uv$
coverage necessary to perform the above tasks.  The point spread
function, or beam, of an array is the Fourier transform of the
$uv$ plane. The beam is easily computed from the coordinates of
the antennas. Unfortunately, there is no analytic solution to the
inverse problem of creating an array configuration for a desired
beam.

The first large two-dimensional radio astronomical array was the
Very Large Array \citep{Thompson:VLA} which has a three-armed
$Y$-shaped configuration.  This configuration was first considered
because it incorporated straight lines of antennas yet also
distributed the antennas over a two dimensional region.  The
ability of such a configuration to cover the $uv$ plane was
supported by empirical studies of the transfer function, and
positions within the $Y$ were chosen through an optimization
procedure.  In more recent work, various procedures for optimizing
the performance of two-dimensional arrays have been developed
\citep{Boone:2001,Cornwell:1988,Keto:1997}.  These studies focused
on array performance as the sole objective, and involved a
relatively small number of antennas.  Radio interferometric arrays
such as the Atacama Large Millimeter
Array\footnote{www.alma.nrao.edu}, the Allen Telescope
Array\footnote{www.seti.org/science/ata.html}, the Square
Kilometer Array\footnote{www.skatelescope.org}, and the Low
Frequency Array (LOFAR)\footnote{www.lofar.org}, take advantage of
advances in signal processing to construct a large aperture from a
large number of relatively small antenna elements, or stations.
For these arrays, the cost of connecting the stations can be a
significant fraction of the total cost.  As part of the design
effort for LOFAR, we have performed an optimization of
two-dimensional configurations with two objectives considered
simultaneously, array performance and cable length.

We have developed models for array performance and cost that
provide metrics used in the optimization.  It is generally
recognized that array performance is improved as the beam sidelobe
level is minimized, and that optimizing the beam of an
instantaneous monochromatic observation centered at zenith allows
for good imaging quality of the array (see for example,
\citet{Boone:2002}, \citet{Cornwell:1988}, \citet{Kogan:1997},
\citet{Woody:1999}). Parseval's theorem implies that the beam
sidelobe level can be minimized by uniform sampling in the $uv$
plane~\citep{Cornwell:1988}. Depending on the scientific goals of
the user, it may be better to optimize on either the beam shape or
the $uv$ distribution. Here we adopt an approach similar to
Cornwell's and construct a metric based on a uniform $uv$
distribution. For the opposing metric of cost, we focus on the
length of the cable required to connect all the stations. As
information on costs associated with cable laying and the
constraints provided by terrain becomes available, the cost metric
can be made more sophisticated.  We describe our models and
simulation in detail in Section~\ref{MSI}.

We have developed a  framework, based on a genetic algorithm, for
rapid exploration of the objective space pertaining to multiple
objectives.  Previous work by \citet{Cornwell:1988} suggests that
the objective space is highly nonlinear with respect to the beam
computation, and that the surfaces representing the performance
metric being optimized can be complex.  Gradient search techniques
are prone to getting trapped in local minima. Thus, the use of
heuristic techniques such as simulated annealing, neural networks,
and genetic algorithms, which are all better at handling nonlinear
objective spaces, need to be used, though, these all require
greater computational resources.  Genetic algorithms have been
used in the past for antenna array design by \citet{Haupt:1994}
and \citet{Yan:1997}, but have focused on compact arrays affected
by antenna coupling effects. They showed that genetic algorithms
are good at solving general array configuration optimization
problems. Here we focus on optimizing for spatially large, sparse
arrays for astronomical imaging performance. Genetic algorithms
also have the advantage that they produce many configurations in a
single run, giving a ``Pareto front'' \citep{Zitzler:2002} of many
optimal solutions at once. In contrast, simulated annealing and
neural networks converge to a single solution, requiring several
simulation runs to determine a Pareto front.   A discussion of our
framework for multiobjective optimization is given in
Section~\ref{FMO}.

Research dealing with the placement of nodes in network design has
been conducted in networking in electrical engineering. From the
early 1990's until recently, a large body of research was devoted
to the Base Station (BS) location problem for cellular phone
networks. At that time the problem was to find the optimal
location of BS (transmitters) in order to satisfactorily cover
subscribers in $xy$-space. Although this problem differs in many
aspects from the radio telescope array problem (notably because
here stations are connected via cables and the goodness of
placement is a strongly non-linear function of all stations in the
array), it is insightful to review the methods used. These range
from Dynamic Programming \citep{Rose:2001}, to Genetic Algorithms
\citep{Han:2001}, \citep{Meunier:2000} and Tabu Search
\citep{Amaldi:2002}. Some of these non-trivial communication
models take into account the limitations imposed by the terrain.

In Section~\ref{Results} we present our results for station
placements and visibility placements, and the explored objective
space for 27-, 60- 100- and 160-station configurations.  The
27-station example allows a comparison with the well-known VLA
configuration; the other simulations are for station numbers
within the range planned for LOFAR. One may ask whether the
optimal configurations found by the multiobjective optimization
framework in Section~\ref{Results} are indeed Pareto optimal, or
whether the results are biased by the use of genetic algorithms.
For this reason array optimization was carried out with simulated
annealing in Section~\ref{SA}. This serves to benchmark and
confirm the configurations found earlier and broadens our
understanding of the performance versus cost tradeoff in telescope
array design. In Section~\ref{Conclusions} we present our
conclusions and our plans for future work.

\section{Model Setup $\&$ Integration}\label{MSI}
\subsection{Design Parameters}\label{Parameters}

Design parameters are quantities that stay fixed during
optimization, but are changed when exploring different designs. We
have three key parameters in our simulations.  The first is simply
the number of stations, $N_{\rm stations}$, which determined the
number of visibility points, $N_{uv}$ according to

\begin{equation}\label{eq_Nuv}
 N_{uv} = (N_{\rm stations})(N_{\rm stations} - 1)
\end{equation}

\noindent (note that because of the Hermitian property of the
visibility function, the number of {\it independent} visibility
points is half this number).  The  number of stations has a great
effect on both the required cable length and the number of $uv$
points in the visibility plane.

The second parameter is the desired radial distribution of the
$uv$ points in the visibility plane.  Parseval's theorem ensures
that minimizing sidelobe levels can be achieved simply by
requiring that no two $uv$ points in the Fourier plane be
redundant \citep{Cornwell:1988}.  Cornwell implemented and
expanded upon this by maximizing the mean distance between $uv$
points as a means to acquire the least redundant $uv$ spacing.  In
this paper we present results for a uniform $uv$ distribution
($\varpropto r^0$), leaving more centrally condensed
concentrations, as is desired for LOFAR, for future work. Figure
\ref{nom_grid} shows a nominal uniform $uv$ distribution for a
27-station configuration. Other radially symmetric $uv$
distributions may be considered as well, i.e. power law or
Gaussian distributions, which are not pictured here.

The third parameter is a size constraint imposed on the placement
of stations.  In accordance with LOFAR site constraints, we have
chosen our terrain to be a circle with diameter of 400 kilometers.
Often large arrays are restricted to a certain piece of land and a
size constraint is necessary in the optimization of the station
placements. As will be seen in Section~\ref{Results}, circular
configurations fill the uv space more than that of Reuleaux
triangle configurations that are size constrained. Unconstrained
simulations may yield different interesting results which may have
larger geometric shapes, such as the Reuleaux triangle, appearing
as the maximum performance configuration. Studies with
unconstrained optimizations will need to be done in future work

\subsection{Design Variables}

Design variables are the quantities that are allowed to vary
during optimization.  The design variables for the optimizations
we carried out are the $xy$ positions of the stations in the
antenna plane.  The number of design variables depends on the
design parameter of number of stations, $N_{stations}$, and is
just twice this number.  The $xy$ positions are varied throughout
the simulation to obtain an optimal set of solutions.

We initialized the $xy$ positions of the stations according to the
well-known topologies shown in Figure \ref{init_seed}, including
\citet{Kogan:1997} circular arrays, \citet{Keto:1997} Reuleaux
triangles, and VLA-like configurations.  We chose these seeds to
search for improvements on already known configurations that
perform well.

In order to explore new regions of the objective space we
initially experimented with seeding our simulations with
non-geometric arrays; for example, arrays of antennas whose
placements were chosen randomly from a uniform distribution in the
$xy$ plane. Optimization attempts initialized exclusively with
non-geometric seeds never successfully evolved to the more
geometric non-dominated configurations shown in
Section~\ref{Results}. In other words, we were more successful in
generating optimal arrays starting from geometric seed solutions
rather than trying to evolve from purely random initial seeds.  An
interesting discussion on the role of randomness and entropy in
array design is provided in Section~\ref{SA}. The difficulty in
finding unique geometric configurations in the large space of
possible antenna designs is the main challenge addressed in this
paper. Sections \ref{Results}, \ref{SA}  \& \ref{Conclusions} will
address this problem further.

\subsection{Design Objectives}\label{objectives}

Design objectives quantify the array designer's desired properties
of the array, and require that metrics be defined that allow the
genetic algorithm to evaluate the fitness of a particular design.
The choice of metrics can be changed to demonstrate trade-offs
between any metrics the user desires.  We have chosen our metrics
based in part upon speed in computing the algorithms as well as
balancing imaging performance and infrastructure cost. It is
important to note that the choice of the algorithms used in
evaluating the metrics is up to the user and the goal that is in
mind.

\subsubsection{Cable Length Minimization}\label{cable}

The first metric we chose is that of the cable length, $L$, which
is to be minimized.  For any configuration of stations, there is
an analytic solution to the minimum cable length problem (also
known as Steiner's problem in graphs \citep{Dreyfus:1971} or
minimum spanning trees \citep{Nesetril:2000}).  We used the Single
Linkage algorithm \citep{Sneath:1957} in the simulations.  In
general, Steiner's problem also addresses the issue of the minimum
{\it cost} of laying cable by associating rules for costing with
the placement of the antennas.  For now, without further
information on the relationship between cable length and cost, we
assume that each unit of cable length is of uniform cost and that
simply minimizing the cable length is sufficient.  Information
providing cable laying costs, or something comparable (i.e.,
site-specific maps) can be used to minimize cost instead of
length.  We validated the cable length minimization code used in
the optimization through comparison to simple geometries and
specific test cases.

\subsubsection{Array Performance}

The second metric we chose is that of the imaging performance of
the array, which is to be maximized.  Cornwell's algorithm
maximized the mean distance between $uv$ points as a way to
eliminate redundant points in the $uv$ plane.   Elegant as this
analysis is, it is very computationally expensive and thus not
well suited for optimizations with as many as the 160 stations
considered here. We have developed a lower-ordered algorithm,
similar to Cornwell's, to calculate the distribution of $uv$
points for a much larger number of stations.  Our method first
calculates a nominal grid for the number of stations.  The nominal
ideal grid consists of $uv$ points placed according to a radially
uniform distribution, and then spaced equally in azimuth at each
radius. At each radius, the azimuthal component of the $uv$ points
in the nominal grid is given a slight random offset to induce
another level of non-redundancy (Figure \ref{nom_grid}). The
nominal grid is calculated once per simulation and used as a
benchmark for the evaluation of each design considered in the
optimization.  Actual $uv$ distributions are calculated from Eqs.
\ref{eq_u} $\&$ \ref{eq_v}, where $x$ and $y$ are the ground
positions of the stations measured in a convenient set of units,
such as kilometers or wavelength:

\begin{eqnarray}
 u_{i,j} &=& {x_i - x_j} \label{eq_u} \\
 v_{i,j} &=& {y_i - y_j} \label{eq_v}
\end{eqnarray}

\noindent where~~$i\ne j$~~$\&$~~$i,j$~$\in\left\{
1,2,\ldots,N_{stations} \right\}$.  An actual $uv$ point from a
design is associated with the nearest $uv$ point in the nominal
grid.  If one or more $uv$ points are associated with a nominal
grid point, the nominal grid point is considered filled.  $N_{\rm
UVactual}$ is the number of nominal grid points which have an
actual $uv$ point associated with them.  The metric is defined by
counting all nominal grid points which have been filled ($N_{\rm
UVactual}$) and subtracting them from the total number of $uv$
points ($N_{UV}$).  This difference is then divided by the total
number of $uv$ points to give the metric $M$, which is the
percentage of nominal grid points that are not filled:

\begin{equation}\label{eq_M}
 M = \frac{N_{UV} - N_{\rm UVactual}}{N_{UV}}
\end{equation}

\noindent This metric always returns a number between zero
(ideally) and one, the latter hypothetically indicating that all
nominal $uv$ points are unfilled. It is desired to have all of the
nominal baselines filled, thus having the array performance
metric, $M$, be zero. Usually this is not physically possible, but
our goal is to sample the Fourier plane as uniformly as possible.
Greater sampling with lower signal-to-noise is desired over a
higher signal-to-noise of fewer Fourier components
\citep{Keto:1997}. Our definition of $M$ ensures that the smaller
the deviation from the nominal case, the better the design appears
in the optimization.

\section{Framework for Multiobjective Optimization}\label{FMO}

\subsection{Genetic Algorithm}\label{GA}

We have implemented a genetic algorithm \citep{Zitzler:2002} with
tournament selection \citep{Pohlheim:1997} for the framework of
the optimizations.  Genetic algorithms are based upon Darwin's
Theory of Evolution (Darwin 1859) using ideas of an evolving
population which improves by combining station placement
information from different configurations (mating), by discarding
poor designs (selection), and by randomly changing the population
to promote diversity (mutation).  We chose to use a genetic
algorithm as the main optimization method for a couple of reasons.
First, the genetic algorithm method, like simulated annealing
\citep{Kirkpatrick:1983} and neural networks \citep{Haykin:1999},
is heuristic, using randomization and statistical techniques
instead of gradient searches.  As we discussed in Section 2, our
metrics are highly nonlinear, and gradient searches are likely to
get trapped in local minima. Second, genetic algorithms are a
population based technique.  This means that they will produce
many optimized configurations simultaneously, which is very
beneficial for multiobjective optimization where a Pareto front of
solutions is desired. They also allow many different initial
guesses for the configurations to be considered at once.

The design of our genetic algorithm involved choices in the
operators of selection, crossover, and mutation.  In the
selection, we wish to filter the population so that the best
members are kept and the worst are discarded.  Many different
types of selection routines have been developed for genetic
algorithms; we have chosen tournament selection (Pohlheim 1997).
Tournament selection takes two different members of the population
(in our case two array configurations) and compares them according
to the metrics. The better of the two designs is retained while
the loser is discarded.  In multiobjective optimization it is
possible that one member is better in one metric while the other
member is better in the second metric; in this case both
configurations are copied once into the next generation.  If one
is better in both metrics, it is coped twice and therefore has a
greater effect on the next generation.  We chose tournament
selection because it requires no weighting of the objectives at
the time of selection. Rather, weighting can be applied after the
optimization and is independent of it, as we discuss further
below. The crossover operator implements the mating part of the
algorithm. In our crossover routine, station $xy$ coordinates are
swapped between stations, making different combinations of array
configurations. One hopes that as the generations evolve, good
configurations will dominate and propagate through the population.
Convergence is reached when the changes in the individual members
of the population become small.  The mutation operator introduces
small random changes in the $xy$ coordinates of the station
placements, allowing an individual station to move to a completely
new place. Mutation is done to keep diversity in the population
and to explore parts of the objective space that might not be
reached by selection and crossover alone. An additional operator,
elitism, was added to the framework to give us control in
emphasizing the population in different parts of the objective
space. Elitism allows us to make copies of any configuration and
insert them into the population, randomly replacing other designs.
We chose to add additional configurations of anchor solutions to
expand the Pareto front near the anchor solutions.

The process of selection, crossover, mutation, and elitism are
repeated, improving the population through generations.  Figure
\ref{flow} shows the flow of our framework.

\subsection{Pareto Optimality}\label{Pareto}

Multiobjective optimization introduces a trade off between the
design objectives, producing an optimal family of solutions which
is a subset of the objective space.  Figure \ref{design_space}
shows a schematic representation of a two-dimensional (two
objective) objective space with important features identified. The
non-dominated solutions are defined as the set of feasible
solutions such that there are no other solutions in the objective
space which improves one design objective without reducing the
optimality in another design objective.  The subset of optimal
solutions is all of the non-dominated solutions which lie on or
near the Pareto front \citep{Zitzler:2002}.  The Pareto front is a
theoretical front made up of non-dominated solutions with full
convergence.  Our results presented in Section~\ref{Results} show
all solutions which we find to be non-dominated.

There are two types of solutions on the Pareto front which are of
special interest:  the ``anchor points'' and the ``nadir-utopia
point.''  The anchor points are the points on the Pareto front
which achieve the best value of one of the design objectives
subject to the constraints and fixed parameterizations of the
problem formulation.  In our optimizations, since there are two
objectives, there are two anchor points.  The nadir-utopia point
is found by first normalizing the axes with respect to the anchor
points located at opposite corners of the normalized objective
space. In the presentation of our simulations both anchor points
have been rescaled to a value of unity.  The ``utopia point,'' or
the theoretical optimum which is not achievable in practice, is at
the corner of the objective space that represents improvement in
both design objectives (the origin in our plots).  The
nadir-utopia point is defined as the point on the Pareto front
which is at the minimum Euclidean distance, $D_{NU}$, from the
utopia point in the normalized coordinate system:

\begin{equation}\label{eq_mindist}
 D_{NU} =
 \min{\sqrt{\left(\frac{J_{1}-J_{1}^{*}}{J_{1norm}}\right)^2+\left(\frac{J_{2}-J_
{2}^{*}}{J_{2norm}}\right)^2}}
\end{equation}

\noindent where $\{J_{1}^{*},J_{2}^{*}\}$ represents the utopia
point, $\{J_{1},J_{2}\}$ represent points on the Pareto front, and
$\{J_{1norm},J_{2norm}\}$ are the normalizations of the axes on
the objective space plot.  Different relative weighting of the
objectives can be implemented by different scalings of the axes,
resulting in different nadir-utopia points.

\section{Results}\label{Results}
\subsection{Simulation Parameters}

We performed optimizations for arrays of 27-, 60-, 100-, and 160-
stations. Table \ref{SimPar} shows the parameters used in these
optimizations: the number of $uv$ points ($N_{UV}$), and the
genetic algorithm parameters of population size, number of
generations, mutation rate, elitism rate, and crossover rate.  The
population is the number of designs which are being operated upon
by the optimizer at one iteration.  As the number of stations
increases, the number of design variables increases accordingly
and  this usually requires a larger population to maintain
diversity throughout the simulation.  As the population increases,
so does the computation time.  The number of generations
determines how many iterations the optimizer runs through and is
set at a  high number to assure convergence in the solution.  The
mutation rate determines the number of random station position
shifts per generation.  This allows different parts of the
objective space to be explored.  The elitism rate determines the
number of anchor points which are reinserted back into the next
generation to aid in expanding the objective space near the anchor
points.  The crossover rate determines how many designs from the
population are mated per generation.  Crossover is the essential
operation in genetic algorithms, as it is used to pass on desired
characteristics throughout an optimization run.

All runs were done on a Pentium 4 2000~MHz machine with 1 gigabyte
of RAM.  Since the algorithms used in the models go as $N^2$, the
times for simulations also increase by roughly $N^2$, making it
much more computationally intensive as the number of stations
increases.  The execution time for each simulation is given in
Table \ref{SimPar}.

\subsection{Configurations}

Figures \ref{27} - \ref{160} show the 27-, 60-, 100-, 160-station
configurations.  Each figure consists of six panels.  The top
three panels in each figure show the $xy$ station placements.  The
stations are circled and cable connections are shown with lines.
 The bottom three panels show the $uv$ coverage corresponding to
the $xy$ station placements directly above them.  From left to
right the configurations presented are those of minimum cable
length, nadir-utopia, and maximum array performance. The minimum
cable length and maximum array performance configurations are the
anchor solutions defined in Section~\ref{Pareto}. The cable length
metric is given in kilometers assuming an overall array diameter
of 400~km (the LOFAR specification), and the array design metric,
$M$ (defined in the figures as $UV Density$), is given as the
fraction of empty nominal $uv$ baselines with respect to the total
number of baselines. Results are presented for an equal weighting
of the objectives. Smaller values are better for both metrics.

The two objectives of minimizing the cable length and achieving
the desired $uv$ distribution oppose each other.  The first tends
to clump stations together to reduce cable length, while the
latter tends to spread stations apart to obtain new $uv$ points.
 Our multiobjective optimization is a way of searching the
configuration space for good trade-off solutions.  The solutions
display a wide Pareto front showing the trade-off between
decreasing cable length, and thus decreasing cost, and improving
array performance.  Along the Pareto front designs range from
VLA-like structures to ring-like structures.  Non-geometric arrays
whose stations are placed randomly are clearly not Pareto-optimal.
This statement is confirmed by simulated annealing in
Section~\ref{SA}.

One expects minimal cable anchor solutions to be highly condensed
designs, but it should be noted that there needs to be a minimum
performance that we, as  array designers wish to consider, or the
minimum cable anchor solution will always be a highly condensed
array with all stations clumping to a point.  Solutions that have
VLA-like configurations are chosen to be the minimum cable
configurations under consideration.  We have chosen the VLA as the
minimum cable configuration because designs with lower cable
length also tend to have VLA-like characteristics, but do not
stretch out to the outer boundaries of the $xy$ plane.

In general, minimum cable solutions appeared as slightly
randomized VLA-like configurations, while best array performance
was achieved by ring-like configurations with inward reaching
arms.  Solutions near the nadir-utopia point consisted of hybrid
solutions of different initial seeds and Reuleaux triangles.

Figures~\ref{27ds} - \ref{160ds} show the 27-, 60-, 100-, and
160-station objective spaces.  The axes of all the objective
spaces are set to the same length so as to facilitate comparisons
between the different objective spaces for the different numbers
of stations. Cable length is on the x-axis, while the $uv$ density
metric, $M$, is on the y-axis.  $UV$ density metric values can
range from close to zero (filled aperture), to nearly one (very
poor $uv$ coverage).  Since the best solutions presented here
asymptote at above 0.25, this is given as a lower boundary on the
axis. Since VLA-like designs were chosen as the minimum cable
configurations ($uv$ density values $\approx$ 0.55), the upper
boundaries of the $uv$ density metric were cut off at an arbitrary
value of 0.75 to show a few designs which achieved lower cable
lengths and lower performance but are still formally Pareto
optimal.

The objective spaces show the initial seeds, the theoretical
Pareto front, the actual non-dominated solutions, and the
evolution of the population over generations.  Initial seed
families were inserted with varying azimuthal or radial
distributions.  This can be seen in the (a) panel of the objective
spaces as a spread in the performance metric for all cases. This
was done to introduce another level of diversity into the initial
population. An interesting, and unexpected result, that can be
seen in the objective spaces is that the population increasingly
evolves away from the Pareto front as the number of stations
increases. This observation is discussed in further detail
throughout the rest of the paper, but is worth noting here. In
these cases, the genetic algorithm framework fills in the gaps
between different initial seeds by hybridization, and also expands
the Pareto front near the anchor solutions.  It does not, however,
expand the Pareto front near the nadir-utopia point for the
metrics that were chosen.

\subsubsection{Performance Comparison to Past Results}

How do genetic algorithm results compare to that of past work?
Work that has been done in the past has mainly focussed on array
performance as the sole objective.  Thus, a fair comparison has to
take into account that we are trying to achieve both high
performance and low cost, as represented by cable length. Our high
performance anchor solutions should compare well to solutions
achieved by other methods with similar performance metrics.  A
good comparison can be made to the configurations done
by~\citet{Keto:1997} and \citet{Boone:2001,Boone:2002}.  Shown in
the third panel of Figures~\ref{27} - \ref{160} are the high
performance arrays. It can be seen that many of them are circular
with inward reaching arms.  The 60-station configurations can be
directly related to configurations for uniform $uv$ distribution
done in \citet{Boone:2001,Boone:2002}.  The Reuleaux triangles
seen in \citet{Keto:1997} do not show up as optimum performance
configurations, but rather as nadir-utopia solutions.  This can be
attributed to the observation that the $uv$ plane is not
completely filled in Figures~\ref{27ds} - \ref{160ds} because the
maximum baseline length for a Reuleaux triangle is smaller than
that of a circle, so the outer most annulus in the $uv$ is
missing. This could be changed if our maximum radius constraint
were not hard bound, allowing the Reuleaux triangle's maximum
baseline to match that of a circle configuration.  In turn, this
would also increase the amount of cable length associated with
that configuration.

\section{Benchmarking with Simulated Annealing}\label{SA}

It is desirable to perform optimization on any design problem with
at least two separate methods. This helps in assessing whether the
converged solutions are truly optimal, or non-dominated as in our
case, or whether they are merely an artifact of the capabilities
of the chosen algorithm. This section presents array optimization
results obtained by Simulated Annealing (SA). After a brief
explanation of the algorithm, results with $N_{stations}=27$ are
used as a benchmark and compared against the earlier results
obtained by genetic multiobjective optimization.

\subsection{Statistical Mechanics and Array Configurations}

The fundamental concept of Simulated Annealing is based on the
Metropolis algorithm \citep{Metropolis:1953} for simulating the
behavior of an ensemble of atoms that are cooled slowly from their
melted state to their low energy ground state. The ground state
corresponds to the global optimum we are seeking in topological
optimization. Simulated Annealing is credited to Kirkpatrick,
Gelatt and Vecchi~\citep{Kirkpatrick:1983} and this article
closely follows their implementation.

In order to apply Simulated Annealing to array configurations, we
must first introduce the notion of ``system energy''. In order to
be consistent with our definition of design objectives in
Section~\ref{objectives}, let

\begin{equation}\label{eq:E}
E(R_i)=E(x,y)= \alpha \cdot \frac{M(x,y)}{M_{avg}} + (1-\alpha)
\cdot \frac{L(x,y)}{L_{avg}}
\end{equation}

be the surrogate for \emph{energy} of a particular array. The
metrics $M$ and $L$ for $uv$ density and cable length were defined
in Equation (\ref{eq_M}) and Section~\ref{cable}, respectively.
$M_{avg}$ and $L_{avg}$ are normalization parameters determined
from a randomly generated array population as discussed below. The
parameter $\alpha \in [0,1]$ can be tuned to emphasize performance
($\alpha=1$) or cable length ($\alpha=0$) in the energy function.

\subsubsection{Statistical Properties of Random Arrays}

In order to understand the average properties of random arrays we
generated 100 arrays, whereby the $x$ and $y$ coordinates are
distributed within a 400 km diameter according to a uniform
probability density. The positions of the random arrays in
objective space are plotted in Figure~\ref{SAobj}. The statistics
of the random population are contained in Table~\ref{tab:SA}. The
normalization parameters are $M_{avg}=0.6413$ and $L_{avg}=1081$
km, respectively. We choose the representative from this
population, which comes closest to the average $uv$ density and
cable length as the initial configuration, $R_o=[x_o, y_o]$. This
initial configuration has a cable length of 1087.2 km and a $uv$
density of 0.6382. Its position in objective space is depicted in
Figure~\ref{SAobj} by a dark square and its topology is
represented in Figure~\ref{SAtop}(a).

\subsection{Simulated Annealing Algorithm}

 The ultimate goal of Simulated Annealing is to find the ground
 state(s), i.e. the minimum energy configuration(s), with a relatively
 small amount of computation. Minimum energy states are those that have a high likelihood of
 existence at low temperature. The likelihood that a
 configuration, $R_i$, is allowed to exist is equal to the Boltzmann probability factor
 \begin{equation}\label{eq:Pri}
    P(R_i)=\exp {\left( - \frac{E \left( R_i
    \right) } {k_B \cdot T} \right)}
\end{equation}

whereby we often set $k_B=1$ for convenience. One can see that, at
the same temperature, $T$, lower energy configurations are more
likely to occur than higher energy configurations. This concept is
at the core of Simulated Annealing.

A block diagram of Simulated Annealing is provided in
Figure~\ref{SAflow}. The algorithm begins with an initial
configuration, $R_o$ and initial temperature $T_o$. This
configuration can be random or an initial best guess. The energy
of the initial configuration, $E(R_o)$ is evaluated. Next, a
perturbed configuration, $R_{i+1}$ is created by (slightly)
modifying the current configuration, $R_i$. For array optimization
a perturbation consists of moving one station to a new, random
location. Next, the energy, $E(R_{i+1})$ and energy difference
$\Delta E=E(R_{i+1})-E(R_i)$ are computed. If $\Delta E <0$, i.e.
the new perturbed configuration is automatically accepted as the
new configuration. If, on the other hand, $\Delta E >0$, we
generate a uniformly distributed random number $\nu \in [0,1]$ and
compare it with the Boltzmann probability $ P ( \Delta E )= \exp [
- \Delta(E) / T ] $. If $\nu$ is smaller than $ P (\Delta E) $ the
perturbed solution is accepted even though it is ``worse'',
otherwise the unperturbed configuration, $R_i$, remains as the
current configuration. Next, we check whether or not thermal
equilibrium has been reached at temperature $T_j$. If thermal
equilibrium has not been reached we go on creating and evaluating
perturbed configurations at the same temperature. If thermal
equilibrium has been reached, i.e. when $n_{eq}$ configuration
changes have been accepted at $T_j$, we reduce the system
temperature by some increment $\Delta T$ and start creating and
evaluating configurations at the new, lower temperature
$T_{j+1}=T_j- \Delta T$. The algorithm terminates, once the system
appears "frozen", i.e. when no new configurations have been
accepted in a large number of attempts. There is no guarantee that
the last configuration is the best, such that one usually keeps in
memory the lowest energy configuration encountered during
\emph{Simulated Annealing}.

\subsection{Simulated Annealing Results}

We now present two different results obtained with Simulated
Annealing. First, we optimized the array, setting the energy
tuning parameter to $\alpha=1.0$. This means that we seek maximum
performance ($M$ as small as possible), regardless of cable
length. The convergence history for this case is shown in
Figure~\ref{SAconverge}. The energy of the initial configuration
is $E_o=0.9796$. This is gradually reduced to
$E^*_{\alpha=1}=0.513$ for the performance-optimal array. It is
interesting to see that Simulated Annealing initially behaves
similarly to random search (up until iteration $\approx 800$) and
transitions to behave more like gradient search as system
temperature is lowered. The best configuration found by Simulated
Annealing with $\alpha=1.0$ is shown in Figure~\ref{SAtop}(b). The
position of this array in the objective space (Figure~\ref{SAobj})
with a $uv$ density of 0.329 and cable length of 1451 km is close,
but slightly offset from the Pareto front computed by the genetic
algorithm. It is noteworthy that the earlier hypothesis that
performance-optimal arrays are ``circles with inward reaching
arms'' is confirmed by Simulated Annealing. An analysis of SA
internal parameters shows that, as temperature decreases
exponentially, entropy also drops sharply towards the end of the
annealing process. Since entropy is the natural logarithm of the
number of unique configurations in the ensemble at a given
temperature step $T_j$, we conclude that Simulated Annealing is
able to reduce entropy, therefore transforming a random initial
array with a high degree of disorder to an ordered (more
geometrical) array with lower entropy. This important point will
be discussed again in Section~\ref{Conclusions}.

What about the case where we want to balance $uv$ density and
cable cost of an array? We set $\alpha=0.5$ and start annealing
the initial configuration, $R_o$, anew. The frozen configuration
in this case can be seen in Figure~\ref{SAtop}(c), its apparent
position in the objective space is shown in Figure~\ref{SAobj}.
The cable length of configuration $R^*_{\alpha=0.5}$ is 691.7 km,
while its $uv$ density is 0.618. This topology is clearly
reminiscent of the ``Y'' configurations used in the VLA and those
that were provided as seed solutions in Figure \ref{init_seed}
(upper left). This again, confirms the types of topologies found
by the genetic algorithm in the short-cable-length regime. We
conclude that Simulated Annealing did not find better or
significantly different arrays than the genetic algorithm and that
this second method therefore corroborates the results discussed
earlier in Section~\ref{Results}. This is at least true for the
values of $\alpha$ that were analyzed. Further tests were
conducted and did not change this result.

We attribute the fact that the solutions obtained by Simulated
Annealing do not lie exactly on the Pareto front to the reduced
computational effort (0.2 hours per run), compared to the genetic
optimizations (Table~\ref{SimPar}). Annealing could be repeated
with a slower cooling schedule and more stringent ``freezing''
criteria, which would produce arrays closer to the Pareto front.
This, however, would be of little value as further runs with
simulated annealing are not likely to change the conclusions of
this section.

\section{Conclusions}\label{Conclusions}
\subsection{Objective Space}

The objective spaces presented above show a wide range of Pareto
optimal designs.  In all cases a concave Pareto front developed
with decreasing marginal returns as designs improved in either
metric.  This leaves the choice up to designers for the type of
array that can be afforded.  If cable length is not an issue in
the array design, then  ring-like solutions with inward reaching
arms are a good choice.  If cable length is a major issue, then
slightly randomized VLA-like configurations may be the best
choice.  If a trade-off is desired, the Reuleaux triangle
configurations, or other hybrid designs near the nadir-utopia
point may be chosen. As more objectives are added to the
multiobjective design problem, different Pareto surfaces may give
new tradeoffs between objectives, particularly if the array should
be grown over time.

As the number of stations increased, Pareto optimal solutions were
more and more like the initial seeds of the population.  Why is
this the case?  The number of $uv$ points in the $uv$ plane is
increasing for a fixed $uv$ plane size.  In essence, the density
of the $uv$ plane is increasing.  As the $uv$ density increases,
the gaps in the $uv$ plane become smaller.  Smaller gap sizes,
along with the large number of $uv$ points makes it difficult for
small perturbations from the highly geometric designs to make a
large difference in the array performance metric.  This creates a
diminishing return on increasing the number of stations to fill
the same $uv$ space.  The highly geometric designs, which have
very good cable length qualities, produce very similar results to
designs that have small random perturbations and longer cable
lengths.  In the 27-station case, there were large gaps in the
$uv$ plane, thus small perturbations and hybridization aided in
improving designs with respect to both metrics.  This result
suggests that a cost trade-off may exist between adding more
stations to a highly geometric design, and moving around a fixed
number of stations to create a better $uv$ coverage, a trade-off
not considered in the work we have done to date. There are other
possible considerations as well, such as surface brightness
sensitivity and the computational power required to combine more
signals from more baselines. New objectives need to be introduced
to take these important considerations into account.

\subsection{Configurations}

An interesting feature of our objective space is that for arrays
with a large number of stations well-known highly geometric
configurations are optimum and occupy particular regions on or
near the Pareto front. There is a progression as we go from
$Y$-configurations, to triangles and Reuleaux triangles, to
circles along the Pareto front.  For the smaller number of
stations we found optimum solutions that are significantly better
than the initial seeds.  For the larger number of stations we were
able to improve the array performance, but we did not find any
designs that simultaneously improved array performance and
shortened cable length; i.e., we could not advance the Pareto
front.  It should be noted, however, that these conclusions are
specific to the two metrics and size constraints we considered.
Introducing new design objectives and relaxing the size
constraints may shift the optimum designs away from geometric
arrays.

In our simulations, configurations improved throughout the
optimization runs, but kept the general shape of the initial
seeds.  Perturbations from ideal geometries and reduction in
unnecessary components of the initial seeds were sufficient to
improve the designs. Why were there no new topologies found?  The
initial seeds into the population are highly geometric.  Highly
geometric arrays have smooth pathways for cable configurations to
follow, thus already having quite a low cable length compared to
very similar designs that have random perturbations from the ideal
geometries and similar array performance.

We discovered the fundamental role that entropy (the degree of
randomness) plays in array optimization and how genetic algorithms
and simulated annealing cope with it in different ways. The
strength of genetic algorithms is to maintain a diverse population
of designs, while continuously advancing the best approximation of
the Pareto front. Genetic algorithms are handicapped, however,
when it comes to reducing entropy in its population of arrays. It
is easy for genetic algorithms to go from highly geometric shapes
to more non-geometric shapes, but statistically it is difficult to
create highly geometric shapes from more non-geometric shapes
during crossover and mutation of station configurations.  Figure
\ref{60randDS} shows the objective space for a 60-station
simulation that was run with only non-geometric initial seeds.  As
can be seen from the figure, the non-geometric initial seeds start
significantly off the Pareto front; just as in the case of
simulated annealing. Figure \ref{RandomvsGeometric} shows the
difference between the Pareto fronts for the 60-station simulation
run with geometric initial seeds vs the 60-station simulation run
with non-geometric initial seeds. As can be seen, both converge to
the best trade-off and high performance solutions similarly. As
one moves along the Pareto front to the highly geometric designs
(up and to the left) the non-geometric initial seed simulation did
not populate that area in the Pareto front. This supports the
entropy argument that it is difficult to produce highly geometric
seeds from non-geometric ones with genetic algorithms.  The
initial seeds into the genetic population are therefore important,
since by infusing geometrical solutions and enforcing some degree
of elitism the entropy of the overall population can be kept low,
while at the same time exploring beneficial randomization.

The $Y$-configuration is probably that which gives the smallest
cable length while spanning a two-dimensional flat surface of a
given size. Including a $Y$-configuration as one of the initial
seeds ensures that the upper left part of the design space is
sampled.

Highly nonlinear objective spaces, such as this one, also pose the
problem that if one good array configuration swaps station
placement information with another good array configuration, the
resulting array is not necessarily going to be an improvement, but
more than likely it will be less optimal if the arrays are very
different from each other. Simulated Annealing, on the other hand,
is very apt at reducing entropy by its very nature. Geometrical
array configurations, such as the minimum-cable-length ``Y''s, can
be found by simulated annealing, see Figure~\ref{SAtop}(c),
starting from random starting points and manual seed solutions are
not required. The problem, however, is that in order to explore
the entire multiobjective space with simulated annealing, the
parameter $\alpha$ must be tuned in many small increments and a
separate optimization must be run for each setting. Even a uniform
sweep of such tuning parameters cannot guarantee that a good
approximation of the Pareto front can always be found. Also, it is
not clear that Simulated Annealing would be computationally less
expensive than the genetic framework once slower cooling schedules
and more conservative freezing criteria are introduced.

This is not to suggest that the genetic framework is perfect as it
stands. Perhaps further improvement can come from implementing
different genetic algorithm techniques. An example is selective
mating, a modification to the mating algorithm which restricts
designs which are too dissimilar from exchanging information.
Also, there is current research in the optimization community
trying to combine the best features of genetic algorithms and
simulated annealing in a new class of hybrid algorithms.

\subsection{Future Work}

We have developed a framework for optimization of antenna arrays
that can be used to address a number of interesting issues in
array design.  There are many considerations in the design of
antenna arrays that we have not yet addressed. The issue of site
constraints is one interesting concept that most ground based
arrays will face.  As array designs come to fruition, they must
ultimately deal with the terrain upon which they will be built.
 Using site masks to eliminate areas where stations can be placed
will put a new constraint upon the design.  Conversely, removing
size constraints, such as the maximum diameter discussed in
Section~\ref{Parameters}, and optimizing may yield interesting
results, e.g. increasing the Reuleaux triangle's size will
increase its radius of curvature, increasing its longest baselines
to compete with those of a circular configuration, while still
performing trade offs between performance and cost. In another
area, it is of course the case that different scientific goals
require different $uv$ distributions. Our ability to modify the
nominal distribution of points in the $uv$ plane will allow us to
address a variety of scientific goals. There is also the issue of
phased deployment of arrays \citep{Takeuchi:2000}.  Many arrays
are not built all at once and then just turned on, rather they are
phased into existence.  A phased deployment may allow particular
scientific questions to be addressed early and may lead to an
array that is more extensible over time.  However, this
requirement would significantly affect array design.  Array
robustness is also an issue of larger arrays. Some stations may be
more critical than others, and repair schedules may become
difficult if many stations are disabled at once.  Analysis of
critical components and failure modes of the array can be useful
in improving the efficiency of array configurations.  Array
optimization will continue to be an important tool in the planning
and construction of future radio telescopes. The framework we have
developed will allow issues, such as the ones described above, to
be addressed in an objective way.

\section*{Acknowledgements}

We thank Drs. Miguel Morales, Divya Oberoi, Colin Lonsdale, and
Roger Cappallo for many helpful discussions.  We would also like
to thank the referee for helpful comments and discussion that led
to improvements in this paper.  This work was supported by grant
AST-0121164 from the National Science Foundation.

\clearpage


\begin{figure}
 \caption{Nominal uniform $uv$ distribution for a 27-station (2 $\ast$ 351 $uv$ point) configuration.}
 \label{nom_grid}
\end{figure}

\begin{figure}
 \caption{Initial population seeds.}
 \label{init_seed}
\end{figure}

\begin{figure}
 \caption{Flow diagram of genetic algorithm framework.}
 \label{flow}
\end{figure}

\begin{figure}
 \caption{Normalized objective space.  Anchor and nadir-utopia points are shown as squares.  The utopia point is shown as a triangle.  Non-dominated solutions are shown as both those that lie on the Pareto front (circles) and those which lie off the front (crosses).  The curving thin line is the outer boundary of the objective space.  All evaluated solutions lie between the Pareto front and the outer boundary.}
 \label{design_space}
\end{figure}

\begin{figure}
 \caption{27-station configurations (top) with corresponding $uv$ coverage (bottom).  Minimum cable configuration (left), nadir-utopia configuration (center), and maximum performance configuration (right).  Smaller values are better for both metrics.}
 \label{27}
\end{figure}

\begin{figure}
 \caption{60-station configurations (top) with corresponding $uv$ coverage (bottom).  Minimum cable configuration (left), nadir-utopia configuration (center), and maximum performance configuration (right).  Smaller values are better for both metrics.}
 \label{60}
\end{figure}

\begin{figure}
 \caption{100-station configurations (top) with corresponding $uv$ coverage (bottom).  Minimum cable configuration (left), nadir-utopia configuration (center), and maximum performance configuration (right).  Smaller values are better for both metrics.}
 \label{100}
\end{figure}

\begin{figure}
 \caption{160-station configurations (top) with corresponding $uv$ coverage (bottom).  Minimum cable configuration (left), nadir-utopia configuration (center), and maximum performance configuration (right).  Smaller values are better for both metrics.}
 \label{160}
\end{figure}

\begin{figure}
 \caption{27-station objective space.  (a) Black dots denote initial seeds of the population and are (from top-left to bottom-right) VLA-like configurations, triangles, Reuleaux triangles, and rings.  The line denotes the Pareto front.  (b) Light dots correspond to the initial 10$\%$ of generations, medium shade dots are the next 30$\%$ of generations, and dark dots are the final 60$\%$.  The non-dominated solutions are enclosed in black squares.  Smaller values are better for both metrics.}
 \label{27ds}
\end{figure}

\begin{figure}
 \caption{60-station objective space.  (a) Black dots denote initial seeds of the population and are (from top-left to bottom-right) VLA-like configurations, triangles, Reuleaux triangles, and rings.  The line denotes the Pareto front.  (b) Light dots correspond to the initial 10$\%$ of generations, medium shade dots are the next 30$\%$ of generations, and dark dots are the final 60$\%$.  The non-dominated solutions are enclosed in black squares.}
 \label{60ds}
\end{figure}

\begin{figure}
 \caption{100-station objective space.  (a) Black dots denote initial seeds of the population and are (from top-left to bottom-right) VLA-like configurations, triangles, Reuleaux triangles, and rings.  The line denotes the Pareto front.  (b) Light dots correspond to the initial 10$\%$ of generations, medium shade dots are the next 30$\%$ of generations, and dark dots are the final 60$\%$.  The non-dominated solutions are enclosed in black squares.}
 \label{100ds}
\end{figure}

\begin{figure}
 \caption{160-station objective space.  (a) Black dots denote initial seeds of the population and are (from top-left to bottom-right) VLA-like configurations, triangles, Reuleaux triangles, and rings.  The line denotes the Pareto front.  (b) Light dots correspond to the initial 10$\%$ of generations, medium shade dots are the next 30$\%$ of generations, and dark dots are the final 60$\%$.  The non-dominated solutions are enclosed in black squares.}
 \label{160ds}
\end{figure}

\begin{figure}
 \caption{Simulated Annealing flow diagram}
 \label{SAflow}
\end{figure}

\begin{figure}
 \caption{Objective space for 27-station case with Simulated Annealing from a random initial array.  Smaller values are better for both metrics.}
 \label{SAobj}
\end{figure}

\begin{figure}
 \caption{(a) Initial random array $R_o$, (b) array optimized by SA ($\alpha=1$), (c) array optimized by SA ($\alpha=0.5$).  Smaller values are better for both metrics.}
 \label{SAtop}
\end{figure}

\begin{figure}
 \caption{Simulated Annealing convergence history for case $\alpha=1.0$}
 \label{SAconverge}
\end{figure}

\begin{figure}
 \caption{60-station non-geometric initial seed objective space.  (a) Black dots denote the non-geometric initial seeds of the population.  The line denotes the Pareto front.  (b) Light dots correspond to the initial 10$\%$ of generations, medium shade dots are the next 30$\%$ of generations, and dark dots are the final 60$\%$.  The non-dominated solutions are enclosed in black squares.  Smaller values are better for both metrics.}
 \label{60randDS}
\end{figure}

\begin{figure}
 \caption{60-station comparison of Pareto fronts for geometric versus non-geometric initial seeds.  Smaller values are better for both metrics.}
 \label{RandomvsGeometric}
\end{figure}

\clearpage

\begin{table}
 \begin{center}
  \begin{tabular}{cccccccc}
   \tableline\tableline
   $N_{stations}$ & $N_{UV}$ & Population & Generations & mrate & erate &
 xrate & time \\
   \tableline
   27 & 2 $\ast$ 351 & 500 & 5000 & 1$\%$ & 1$\%$ & 90$\%$ & 10.1 hrs \\
   60 & 2 $\ast$ 1770 & 200 & 5000 & 1$\%$ & 1$\%$ & 90$\%$ & 18.3 hrs \\
   60 non-geometric & 2 $\ast$ 1770 & 200 & 5000 & 1$\%$ & 1$\%$ & 90$\%$ & 24.7 hrs \\
   100 & 2 $\ast$ 4950 & 300 & 6000 & 1$\%$ & 1$\%$ & 90$\%$ & 117.2 hrs \\
   160 & 2 $\ast$ 12720 & 200 & 2000 & 1$\%$ & 1$\%$ & 90$\%$ & 72.3 hrs \\
   \tableline
  \end{tabular}
  \caption{Simulation parameters for genetic algorithm optimization based
 upon varying numbers of stations ($N_{stations}$).  The number of stations
 ($N_{stations}$), the corresponding number of $uv$ points ($N_{UV}$),
 and the
 genetic algorithm parameters of $Population$ and $Generations$ are given.  $N_{UV}$ is given as two times the number of independent $uv$
 points because of the Hermitian property of the visibility function.
 The genetic algorithm parameters of mutation rate ($mrate$), elitism rate
 ($erate$), and crossover rate ($xrate$) are given in percentages.  Simulation
 run times are given in hours.}
  \label{SimPar}
 \end{center}
\end{table}

\begin{table}
  \centering
  \begin{tabular}{ccccc}
    \hline \hline
    parameter & 100 random arrays & Initial Array $R_o$ & Array ($R^*_{\alpha=1.0}$) & Array ($R^*_{\alpha=0.5}$) \\
    \hline
    $M_{avg}$ & 0.6413 & 0.6382 & 0.3290 & 0.6182 \\
    $\sigma (M)$ & 0.0483 & - & - & - \\
    $L_{avg} \textrm{ km}$ & 1081 & 1087.2 & 1451.1 & 691.7 \\
    $\sigma (L) \textrm{ km}$ & 117.3 & - & - & - \\
    \hline
  \end{tabular}
  \caption{Characteristics of random and SA-optimized arrays with $N_{stations}=27$.}\label{tab:SA}
\end{table}

\end{document}